\documentclass[runningheads]{llncs}

\usepackage{graphicx}
\usepackage{booktabs}
\usepackage{tabularx}
\usepackage{hyperref}
\usepackage{xcolor}
\usepackage{microtype}

\begin{document}

\title{What Do AI Agents Actually Change? \\
       An Empirical Taxonomy of Mutation Patterns \\
       in Performance-Improving Pull Requests}
\titlerunning{Mutation Patterns in AI-Agent Performance PRs}

\author{Illia Dovhoshliubnyi\inst{1} \and
        Nima Soroush\inst{1} \and
        Ashkan Sami\inst{1} \and
        Alexander Brownlee\inst{2}}
\authorrunning{I. Dovhoshliubnyi et al.}
\institute{Edinburgh Napier University, Edinburgh, UK \\
\email{illiadovho@gmail.com, \{N.Soroush, A.Sami\}@napier.ac.uk}
\and
University of Stirling, Stirling, UK \\
\email{alexander.brownlee@stir.ac.uk}}
\maketitle

%--- ABSTRACT ------------------------------------------------------------------
\begin{abstract}

AI coding agents are black boxes: we cannot inspect how they generate code,
but we can inspect what they change. This distinction matters for
search-based software engineering (SBSE), where techniques such as genetic
improvement (in the performance-optimisation application we study)
depend on mutation operators that reflect how code is actually transformed.
Fewer than 1\% of the 33{,}596 agent PRs in AIDev-pop target performance,
making each case a rare window into otherwise opaque agent behaviour. We
classify 1{,}254 performance-relevant diff hunks from 216 of these PRs,
spanning five agent systems, against the 18-category syntactic mutation
taxonomy of Even-Mendoza et al.\ (2025) using a dual-LLM intersection
pipeline. Three categories dominate: \emph{name\_modification} (37.0\%),
\emph{object\_creation} (26.4\%), and \emph{type\_change} (22.7\%), a
profile markedly different from prior GI corpora where \emph{no\_change}
accounted for 84\%. Each agent's deployed system commits to a distinctive mutation vocabulary,
and each performance strategy activates a largely disjoint category subset.
Agent identity and target strategy are therefore informative priors that
narrow the effective SBSE operator space.

\textbf{Replication package: 
\url{https://github.com/5uper6rain/ssbse-challenge-2026}}

\end{abstract}


%--- 1. INTRODUCTION -----------------------------------------------------------
\section{Introduction}
\label{sec:intro}
%%%%%%%%%%%%%%%%%%%%%

AI coding agents such as Devin, GitHub Copilot, Cursor, OpenAI Codex, and
Claude Code autonomously submit pull requests to production repositories,
but their mechanisms for deciding what to change are opaque. We cannot
inspect those mechanisms, but we can inspect the outputs: the actual code
transformations they commit.

This distinction matters for SBSE.
Genetic improvement and related techniques depend on mutation operators
grounded in empirical evidence of how code is actually
transformed~\cite{harman2001sbse,jia2011mutation,petke2023program}.
We focus specifically on the performance-improvement application of GI.
Classical mutation taxonomies were derived from human-written patches.
As SBSE is increasingly applied to agent-assisted workflows, an empirical
map of how agents actually transform code is a missing prior:
descriptive of agent behaviour for the purpose of scoping operator
selection in agent-aware tooling, not prescriptive of agent behaviour as
an optimum.

Performance-improving PRs are rare: only 324 of the 33{,}596 PRs in
AIDev-pop~\cite{aidev} carry a performance label ($<$1\%)~\cite{opu2025agentic,peng2026agents},
as most agent PRs target bugs or features. These rare cases are the
instances where agents intentionally optimised code, making post-hoc
mutation analysis directly informative for SBSE.

We address two research questions:

\noindent\textbf{RQ1.} \emph{What syntactic mutation patterns characterise
successful performance PRs from AI coding agents, and how do they differ
from those observed in prior genetic improvement work?}

\noindent\textbf{RQ2.} \emph{Do mutation patterns vary systematically
across agent systems and across performance strategies?}

\section{Dataset and Methodology}
\label{sec:method}

\subsection{Dataset}
We use the AIDev-pop subset of the AIDev dataset~\cite{aidev}: PRs from
five AI coding agents (Devin, GitHub Copilot, Cursor, OpenAI Codex, and
Claude Code) against 100 starred repositories. Of the 324 PRs carrying a
performance label~\cite{peng2026agents}, 280 had a retrievable diff; 269
contained at least one source-code hunk after discarding tests,
configuration, documentation, and lock files. Each PR's diff is decomposed
into contiguous change \emph{hunks}. A first-pass LLM filter
(\texttt{claude-sonnet-4-6}) retains only performance-relevant hunks (those
plausibly affecting runtime, memory, or I/O); cosmetic-only edits and
single-model false positives are discarded by the dual-LLM intersection
(Sec.~\ref{subsec:classification}). After agreement filtering, 216 PRs
contribute at least one accepted hunk, yielding 1{,}254 hunks for analysis.
Each PR is assigned to one of nine optimisation patterns~\cite{peng2026agents}.

%

%We started with 324 agent PRs identified in~\cite{peng2026agents} that carry a performance sub-pattern label out of 33{,}596 PRs in the AIDev-pop a subset of the AIDev dataset ~\cite{aidev}. AI-Dev contanins PRs from Devin, GitHub Copilot, Cursor, OpenAI Codex, and Claude Code.  Of the 324 agent PRs, \textbf{280} carry a performance sub-pattern label; 11 contained no business-logic source files after filtering tests, configuration, and lock files, leaving \textbf{269 PRs} as the working corpus. Of these, \textbf{216 contribute at least one performance-relevant hunk} after filtering, forming the analysis population of
%\textbf{1,254 hunks}. Each labelled PR is assigned to one of nine high-level optimization patterns~\cite{peng2026agents} used in Section~\ref{subsec:patterns}. 

%Each PR's unified diff is decomposed into contiguous change \emph{hunks} and filtered to business-logic source files, discarding tests, configuration, documentation, and lock files. A first-pass LLM filter then identifies performance-relevant hunks---changes plausibly affecting runtime, memory, or I/O behaviour---discarding purely cosmetic edits. Dual-LLM agreement filtering (Section~\ref{subsec:agreement}) ultimately retains \textbf{1,254 hunks} across 216 PRs.

%%%%%
\subsection{Mutation Taxonomy}
\label{subsec:taxonomy}

We adopt the 18-category syntactic mutation taxonomy
of~\cite{evenmendoza2025}, originally derived from manual analysis of
LLM-generated patches on Java projects in a Genetic Improvement (GI)
pipeline. The categories cover structural transformations (e.g.,
\emph{control\_flow}, \emph{statement\_splitting}), identifier-level edits
(\emph{name\_modification}, \emph{type\_change}), and resource-management
changes (\emph{synchronization}, \emph{object\_creation}), plus two
sentinel categories.

Three researchers calibrated category boundaries through six iterative
prompt-design rounds on a held-out 40-hunk sample
(Section~\ref{subsec:classification}; full protocol in replication
package), most notably extending \emph{object\_creation} (cat.~14) to
cover import-statement changes, since importing a new library constitutes a
module-level creation in the performance context.

%%%%%
% \subsection{Mutation Taxonomy}
% \label{subsec:taxonomy}
% We adopt the 18-category syntactic mutation taxonomy of~\cite{evenmendoza2025}, originally derived from manual analysis of
% LLM-generated patches on Java projects in a Genetic Improvement (GI) pipeline. The categories cover structural
% transformations (e.g., \emph{control\_flow}, \emph{statement\_splitting},
% \emph{inlining}), identifier-level edits (\emph{name\_modification},
% \emph{type\_change}), and resource-management (\emph{synchronization},
% \emph{object\_creation}), plus two sentinel categories.
% Table~\ref{tab:taxonomy} reports corpus frequencies.

% Three researchers calibrated category boundaries over two rounds of reviews. \emph{object\_creation} (cat.~14) was extended to cover import-statement changes (importing a library is a new module creation). We added definition for \emph{inlining} (cat.~8), disambiguated \emph{type\_change} (cat.~7) from same-API library substitutions, and narrowed \emph{arithmetic\_manipulation} (cat.~16) to explicit operator changes.
%Three researchers calibrated category boundaries over six rounds. The key adaptation extended \emph{object\_creation} (cat.~14) to cover import-statement changes (importing a library is a new module creation). We also tightened \emph{inlining} (cat.~8), disambiguated \emph{type\_change} (cat.~7) from same-API library substitutions, and narrowed \emph{arithmetic\_manipulation} (cat.~16) to explicit operator changes.

\begin{table}[t]
\caption{Mutation taxonomy with corpus frequencies (multi-label; \% of 1,254 classified hunks).}
\label{tab:taxonomy}
\centering\footnotesize
\begin{tabularx}{\linewidth}{clrr>{\raggedright\arraybackslash}X}
\toprule
\textbf{ID} & \textbf{Category} & \textbf{Count} & \textbf{\%} & \textbf{Description} \\
\midrule
12 & name\_modification        & 463 & 37.0 & Modified variable, function, class name \\
14 & object\_creation          & 330 & 26.4 & Modified object/primitive creation or initialization; new imports \\
7  & type\_change              & 284 & 22.7 & Changed data types or type usage  \\
13 & control\_flow             & 262 & 20.9 & Modified control flow: if/else, loops, switch \\
15 & statement\_splitting      & 232 & 18.5 & Split a statement into multiple lines \\
2  & comment\_modification     & 167 & 13.3 & Modified a comment (add/remove/edit) \\
3  & deleted\_blocks           &  90 &  7.2 & Deleted blocks in a method (all/most/some) \\
16 & arithmetic\_manipulation  &  80 &  6.4 & Boolean var.\ or expr.\ manipulations \\
5  & return\_statement\_mods   &  56 &  4.5 &   Add/remove/edit return statements\\
6  & method\_name\_change      &  24 &  1.9 & Changes to method names \\
9  & exception\_handling       &  17 &  1.4 & Unreachable or reachable exception \\
0  & added\_code\_from\_github &  16 &  1.3 & Added (some arbitrary) code from GitHub \\
11 & synchronization           &  10 &  0.8 & Added synchronization logic \\
8  & inlining                  &   2 &  0.2 & Replacing a function call with actual body \\
17 & dead\_code                &   1 &  0.1 & Added dead code \\
4  & duplicate\_code           &   1 &  0.1 & Duplicate code \\
1  & no\_change                &   0 &  0.0 & No meaningful code change \\
10 & extra\_brackets           &   0 &  0.0 & Added extra brackets \\
\bottomrule
\end{tabularx}
\end{table}

% \subsection{Dual-LLM Classification}
% \label{subsec:classification}
% Each hunk is independently classified by two LLMs (\texttt{claude-sonnet-4-6}
% and \texttt{gpt-5.4}) using the same structured prompt specifying all 18
% categories with definitions, examples, and disambiguation notes.
% Both models were called at API-default temperature~(1.0), \texttt{max\_tokens}
% 256/512, confidence threshold~0.8; total cost \$12 (Claude) +
% \$10 (GPT-5.4).
% Classifier accuracy was established through six calibration rounds on a 40-hunk
% held-out set, achieving \textbf{67.5\% exact accuracy}, 22.5\% partial match,
% and 10\% incorrect.

% \subsection{Agreement Resolution}
% \label{subsec:agreement}
% For each hunk, the two model outputs are compared by category-set intersection:
% \textbf{full agreement} (48.4\%, $n=607$) adopts the label directly;
% \textbf{partial agreement} (51.6\%, $n=647$) adopts the intersection, retaining
% only categories both models assigned; and \textbf{full disagreement} hunks are
% discarded as too ambiguous. Category counts are therefore conservative lower bounds.
\subsection{LLM-as-a-Judge Classification}
\label{subsec:classification}

We adopt an LLM-as-a-judge approach~\cite{zheng2023judging}: each hunk is
independently classified by \texttt{claude-sonnet-4-6} and \texttt{gpt-5.4}
using the same structured prompt (18 categories with definitions, examples,
and disambiguation notes; cost \$22). Prompt accuracy was validated over six
design rounds on a 40-hunk sample (uniform random, seeds 42$\to$777, one
boilerplate repository excluded); two independent runs confirmed 67.5\%
exact and 22.5\% partial accuracy---substantially above chance for 18-class
multilabel. For each hunk the two outputs are compared by category-set
intersection: full agreement (48.4\%, $n{=}607$) adopts the label; partial
agreement (51.6\%, $n{=}647$) takes the intersection; full disagreements
are discarded. Category counts are therefore conservative lower bounds.

%--- 3. RESULTS ----------------------------------------------------------------
\section{Results}
\label{sec:results}

\subsection{RQ1: Category Distribution}
\label{subsec:distribution}

Table~\ref{tab:taxonomy} shows the distribution across classified hunks
(multi-label). \emph{Name\_modification} (37.0\%) leads strongly, followed
by \emph{object\_creation} (26.4\%), \emph{type\_change} (22.7\%),
\emph{control\_flow} (20.9\%), and \emph{statement\_splitting} (18.5\%).
Apriori mining confirms a
\emph{name\_modification}+\emph{type\_change} co-occurrence cluster
(lift~3.07).

The dominance of \emph{name\_modification} warrants interpretation. In 78\%
of \emph{name\_modification} cases (Devin PRs following the
\emph{Performance-Optimized Dependency Selection} pattern), optimisation
manifests as swapping library imports for faster alternatives
(e.g., \texttt{chalk}~$\rightarrow$~\texttt{picocolors}). This makes
\emph{name\_modification} a \emph{carrier category} whose sub-types
(import additions, call-to-cached-variable substitutions, API-level
substitutions) warrant separation in future taxonomy revisions.
Runtime-risk categories---\emph{synchronization} (0.8\%),
\emph{inlining} (0.2\%), \emph{exception\_handling} (1.4\%)---are
seldom employed as performance optimisations.

\paragraph{Comparison with Even-Mendoza et al.}
In the original GI corpus, \emph{no\_change} dominated at 84\% of patches
and \emph{type\_change} showed the best test-passing
rate~\cite{evenmendoza2025}. Our corpus is markedly different:
\emph{no\_change} is entirely absent and \emph{arithmetic\_manipulation}
is rare (6.4\%). GI patches are broadly sampled, whereas AI-agent
performance PRs are goal-directed commits, so taxonomies calibrated on GI
corpora do not transfer directly to agent-generated code.

\subsection{RQ2: Variation by Performance Pattern and Agent}
\label{subsec:variation}

\begin{figure}[t]
  \centering
  \includegraphics[width=\linewidth]{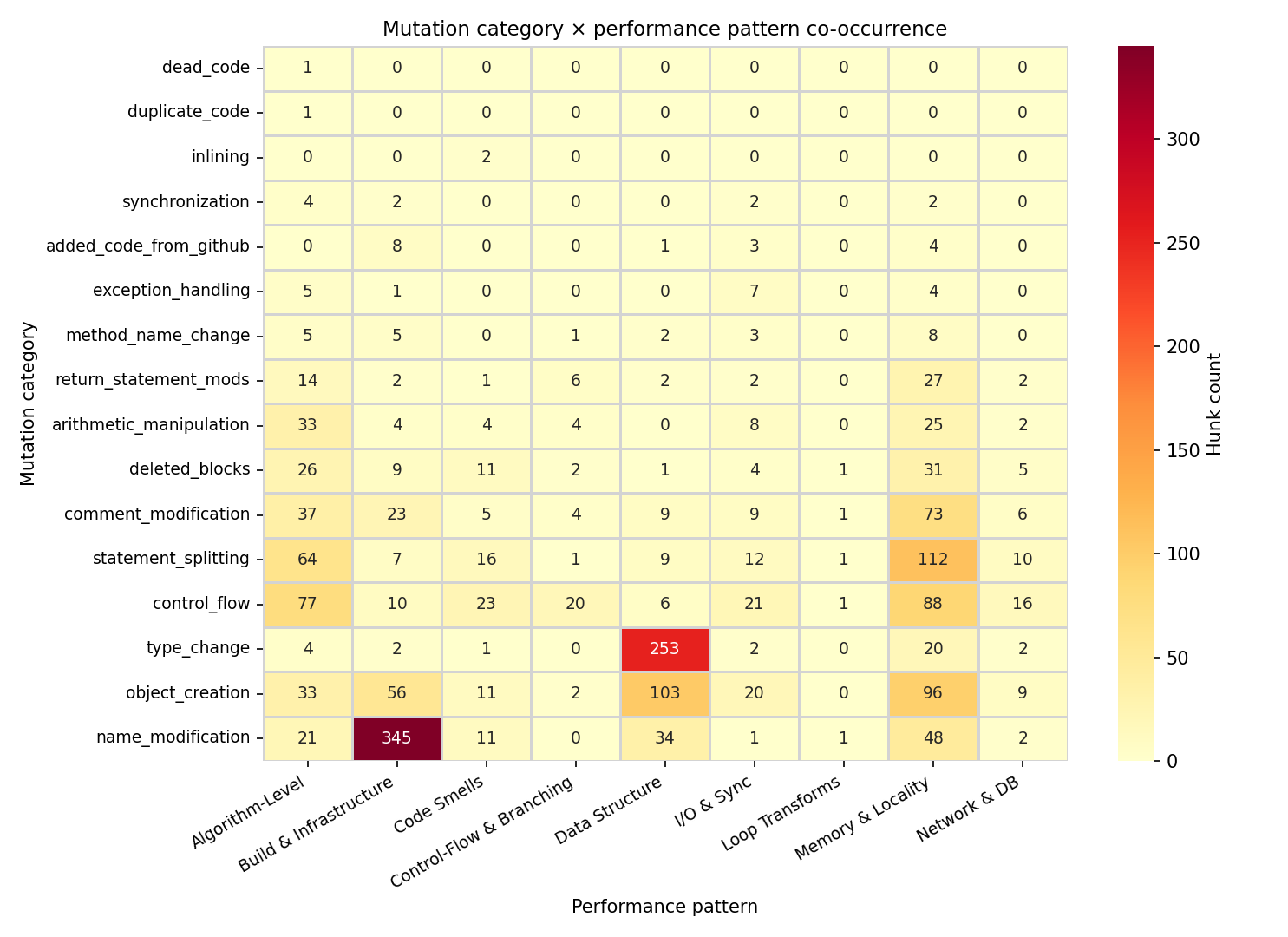}
  \caption{Co-occurrence of all non-zero mutation categories with AIDev
           performance patterns. Cells show hunk-category assignment counts
           (multi-label).}
  \label{fig:heatmap}
\end{figure}

\paragraph{Performance patterns.}
Figure~\ref{fig:heatmap} maps mutation categories to AIDev performance
patterns. Two associations dominate: \emph{type\_change} concentrates in
\emph{Data Structure} (253/284; 89\%) and \emph{name\_modification} in
\emph{Build \& Infrastructure} (345/463; 75\%). \emph{Object\_creation}
bridges \emph{Data Structure} (103) and \emph{Memory \& Locality} (96);
\emph{control\_flow} co-occurs with \emph{object\_creation} in 36.6\% of
PRs. Within each agent's primary ecosystem (Devin/TypeScript $n{=}461$;
Copilot/Rust $n{=}219$; Codex/Go $n{=}120$) one category dominates:
\emph{name\_modification} 77\%, \emph{type\_change} 95\%,
\emph{control\_flow} 63\%. Cross-language stability is not established
(Sec.~\ref{sec:threats}); each agent's profile co-varies with the
language mix it targets.

%%%%%%%%%%%%%%%%%%%%%%%%%%%%%%%%%%%%%%%%
%
%%%%%%%%%%%%%%%%%%%%
\paragraph{Agent profiles.}
Table~\ref{tab:agents} provides a second dimension of variation.
Category assignments are broken down by agent (Claude Code excluded due
to small sample size, $n{=}16$). Four distinct profiles emerge.
\emph{Devin} (670 total) is dominated by \emph{name\_modification} (361;
54\%), reflecting dependency substitution at scale. \emph{GitHub Copilot}
(591) is led by \emph{type\_change} (224; 38\%), consistent with
data-structure upgrades. \emph{OpenAI Codex} (574) shows the broadest
profile: \emph{control\_flow} (136; 24\%) and \emph{object\_creation}
(101; 18\%) lead. \emph{Cursor} (184; smallest sample) shows elevated
\emph{comment\_modification} (41; 22\%); since the perf-relevance filter
excludes cosmetic-only hunks, this reflects annotations co-occurring with
structural edits rather than standalone documentation changes.

Knowing which agent produced the code is thus a complementary, partly
correlated prior alongside target strategy for narrowing the operator space.
%%%%%%%%%%%%%%%%%%%%%
% Each performance strategy activates a characteristic mutation subset: operators
% should be \emph{conditioned on the target performance strategy} rather than drawn
% uniformly from the 18-category space.

% \subsection{Per-Agent Variation}
% \label{subsec:agents}
% Table~\ref{tab:agents} breaks category assignments down by agent (Claude Code
% excluded; $n=16$ assignments). Four distinct profiles emerge.
% \emph{Devin} (670 total) is dominated by \emph{name\_modification} (361; 54\%),
% reflecting dependency substitution at scale.
% \emph{GitHub Copilot} (591) is led by \emph{type\_change} (224; 38\%),
% consistent with data-structure upgrades.
% \emph{OpenAI Codex} (574) shows the broadest structural profile:
% \emph{control\_flow} (136; 24\%) and \emph{object\_creation} (101; 18\%) lead.
% \emph{Cursor} (184) has elevated \emph{comment\_modification} (41; 22\%),
% suggesting it accompanies structural changes with explanatory annotations.
\begin{table}[t]
\caption{Category assignments per agent (top 8 categories; sorted by agent total).}
\label{tab:agents}
\centering\footnotesize
\begin{tabular}{lrrrr}
\toprule
\textbf{Category} & \textbf{Devin} & \textbf{Copilot} & \textbf{Codex} & \textbf{Cursor} \\
\midrule
name\_modification    & 361 &  30 &  61 & 11 \\
object\_creation      &  81 & 116 & 101 & 29 \\
type\_change          &  35 & 224 &  22 &  2 \\
control\_flow         &  51 &  48 & 136 & 24 \\
statement\_splitting  &  65 &  49 &  95 & 21 \\
comment\_modification &  22 &  64 &  37 & 41 \\
deleted\_blocks       &  13 &  17 &  43 & 15 \\
arithmetic\_manip     &  16 &  16 &  29 & 17 \\
\midrule
\textit{All categories (total)} & \textit{670} & \textit{591} & \textit{574} & \textit{184} \\
\bottomrule
\end{tabular}
\end{table}

%--- 4. THREATS TO VALIDITY ----------------------------------------------------
\section{Threats to Validity}
\label{sec:threats}
\textbf{Construct.} The taxonomy was derived from Java LLM-generated patches and
may not capture every mutation type in our multilanguage corpus. We
additionally extended \emph{object\_creation} to include module-level
imports, so its 26.4\% rate exceeds the original Java-method scope and is
not directly comparable to prior corpus statistics. Category boundaries
were calibrated for performance PRs and may not generalise to bug-fix or
feature PRs.
\textbf{Internal.} LLM accuracy is bounded at 67.5\% after six calibration
rounds. Both the first-pass perf-relevance filter and the dual-LLM
intersection err toward exclusion, so reported category counts are
conservative lower bounds reflecting potential LLM
bias~\cite{zheng2023judging}; per-category proportions may still be
inflated for mutations the filter detects most reliably (e.g., explicit
import or rename edits).
\textbf{External.} Performance labels are content-based: merged
($n{=}324$ hunks) and unmerged ($n{=}927$ hunks) sub-corpora overlap on
4/5 top categories (Spearman $\rho{=}0.90$) with shifted emphases---
\emph{control\_flow} and \emph{statement\_splitting} over-represent
merged hunks (${\sim}3{\times}$); \emph{name\_modification} and
\emph{type\_change} over-represent unmerged hunks (Cliff's
$|\delta|\leq0.28$); we retain both. All percentages are
hunk-weighted, so PRs producing many hunks contribute proportionally
more weight than concise PRs. Performance labels themselves originate
from~\cite{peng2026agents}; upstream mislabelling would propagate to
our scope. Agent and language ecosystem are correlated in AIDev-pop
(Devin→TypeScript; Copilot→Rust; Codex→Go), so per-agent profiles
describe deployed systems and may not generalise to other agents or
repositories. Agent and judge-LLM versions are snapshots; results
may shift as these systems evolve.

%--- 5. CONCLUSION -------------------------------------------------------------
\section{Conclusion}
\label{sec:conclusion}
AI coding agents are black boxes, but their outputs are not. By classifying
1{,}254 performance-relevant diff hunks we find agent performance PRs
dominated by \emph{name\_modification}, \emph{object\_creation}, and
\emph{type\_change}~--- sharply different from prior GI corpora (RQ1)~---
with each agent and strategy activating a distinct subset (RQ2;
agent profiles co-vary with the language mix each system targets,
Sec.~\ref{sec:threats}).
In a GI loop conditioned on (target strategy, agent), per-context priors
from Tables~\ref{tab:taxonomy}--\ref{tab:agents} narrow the operator space
from 18 categories to roughly five.

\bibliographystyle{splncs04}
\bibliography{refs}

\end{document}